\documentstyle[graphicx]{europhys}

\def\And{{\rm and\ }}

\def\stars{\bigskip\centerline{***}\medskip}

\newif\ifboo \boofalse

\def\Review#1{\boofalse{\it #1},}
\def\Name#1{{\sc #1},}
\def\Vol#1{\ifboo Vol. {\bf #1}\else{\bf #1}\fi}
\def\Year#1{\ifboo #1\else(#1)\fi}

\def\Page#1{\ifboo {\rm p. #1}\else{\rm #1}\fi}

\def\eurologo{{\rm EURO-\LaTeX\ (does not imply 
endorsement by \sc{Europhysics Letters})}}

\makeatletter
\def\vereq#1#2{\lower3pt\vbox{\baselineskip1.5pt \lineskip1.5pt
\ialign{$\m@th#1\hfill##\hfil$\crcr#2\crcr\sim\crcr}}}
\def\agt{\mathrel{\mathpalette\vereq>}}
\def\alt{\mathrel{\mathpalette\vereq<}}
\makeatother

\newcommand{\kB}{k_{{\rm B}}}
\newcommand{\kT}{\kB T}
\newcommand{\fid}{f^{{\rm (id)}}}
\newcommand{\fex}{f^{{\rm (ex)}}}
\newcommand{\A}{{\mathbf A}}
\newcommand{\B}{{\mathbf B}}
\newcommand{\pav}[1]{\langle #1\rangle_p}
\newcommand{\myphi}{\phi}
\newcommand{\myetal}{{\it et al}}
\newcommand{\eqref}[1]{(\ref{#1})}

\begin{document}
\euro{}{}{}{}
\Date{}
\shorttitle{P. B. WARREN: FLUID-FLUID PHASE SEPARATION ETC.}
\title{Fluid-fluid phase separation in hard spheres\\
with a bimodal size distribution}
\author{P. B. Warren}
\institute{Unilever Research Port Sunlight Laboratory,\\
Quarry Road East, Bebington, Wirral, L63 3JW, UK.}
\rec{}{}
\pacs{
\Pacs{05}{20$-$y}{Statistical mechanics}
\Pacs{64}{75$+$g}{Solubility, segregation and mixing}
\Pacs{82}{70Dd}{Colloids}}
\maketitle

\begin{abstract}
The effect of polydispersity on the phase behaviour of hard spheres is
examined using a moment projection method. It is found that the
Boublik-Mansoori-Carnahan-Starling-Leland equation of state shows a
spinodal instability for a bimodal distribution if the large spheres are
sufficiently polydisperse, and if there is sufficient disparity in mean
size between the small and large spheres. The spinodal instability
direction points to the appearance of a very dense phase of large
spheres.
\end{abstract}

The phase behaviour of hard sphere mixtures has received much attention
recently following the observation by Biben and Hansen \cite{BH} that,
with an improved closure approximation, integral equation theory
indicates the presence of a fluid-fluid spinodal instability in a
bidisperse mixture if the ratio of the large to small sphere diameters
is $\agt5$. Several experimental groups report confirmatory evidence
\cite{Expts,SMS,DYP}. The dense phase of the large spheres is sometimes
observed to be an ordered phase and at other times to be an amorphous
colloidal glass, although some discrepancies in detail remain
\cite{SMS,DYP}.

Numerical solutions to integral equations frequently only suggest the
existence of a spinodal instability, from the apparent divergence of the
structure factor for instance. Thus for binary hard spheres, Caccamo and
Pellicane \cite{CP} can argue that the integral equation evidence
supports fluid-solid segregation rather than fluid-fluid separation.
Recent computer simulations support this scenario, by establishing that
a region of fluid-fluid coexistence exists, but is metastable with
respect to fluid-solid coexistence \cite{DRE}.

Analytic solutions to integral equations, although rare and generally
perceived to be less accurate than numerical solutions, do not suffer
from the same problems. Most notable of these for hard spheres is the
Percus-Yevick closure \cite{Lebowitz}. As reported by Lebowitz and
Rowlinson \cite{LR} though, this closure predicts that a binary hard
sphere fluid is stable for all size ratios and physically accessible
compositions.

The appeal of the Percus-Yevick approximation is tempered by an internal
inconsistency between the compressibility equation of state (EOS) and
the virial EOS. It is precisely this inconsistency that the new closure
approximations mentioned above are designed to cure. A phenomenological
way around this problem though is to interpolate between the two EOS's.
For monodisperse hard spheres, a suitable interpolation leads to the
highly succesful Carnahan and Starling EOS \cite{CS}. For an arbitrary
mixture, the same interpolation leads to an EOS first considered by
Boublik and Mansoori \myetal\ (BMCSL EOS) \cite{BMCSL}. This improved
EOS also predicts phase stability for a binary hard sphere fluid for any
size ratio and composition.

It is well known though that polydispersity enhances phase instability.
An interesting and experimentally relevant question therefore concerns
the effects of size polydispersity on the phase behaviour of hard sphere
mixtures. Fortunately the above EOS's have been extended to polydisperse
systems, for instance the Helmholtz free energy and related
thermodynamic quantities corresponding to the BMCSL EOS are given by
Salacuse and Stell \cite{SS} in terms of the number density and the
first three moments of the size distribution.

For the Percus-Yevick compressibility EOS, Vrij \cite{Vrij} established
that an arbitrarily polydisperse hard sphere fluid is stable, thus
confirming a conjecture of Lebowitz and Rowlinson \cite{LR}. However the
question of whether, under the BMCSL EOS, a binary mixture becomes
unstable if one or both of the species is allowed to become polydisperse
has not to my knowledge been previously investigated. Perhaps
surprisingly in the light of Vrij's result the answer to this question
is ``yes''. Moreover the instability reflects the experimental
observations, albeit at a much larger degree of polydispersity and size
ratio than are seen in reality. Very recently Cuesta has found a similar
spinodal instability in hard spheres with a unimodal, log-normal size
distribution \cite{Cuesta}.

I address the spinodal stability problem using a novel moment projection
method developed recently by Sollich and Cates \cite{SC} and myself
\cite{Warren}. The approach reduces the free energy to one which depends
only on moment densities (defined below). It gives exact results for
cloud and shadow curves (the polydisperse analogues of the binodal) and
for spinodal curves, provided the excess free energy is a function only
of the corresponding moments of the size distribution. Since the BMCSL
excess free energy is precisely of this form, the new tool may be aplied
to this situation. By way of contrast, Cuesta approaches the spinodal
problem by casting it as an integral equation \cite{Cuesta}. He also
finds that if the excess free energy depends only on a few moments, the
integral equation is reducible to a matrix problem, which can in fact be
proved to be identical to the results obtained below. Such a reduction
in dimensionality for the spinodal problem has been noted several times
in the past \cite{Moms}.
 
I outline the approach in general first to indicate its application to
spinodal curves, which was not previously covered in detail. Suppose
that the excess free energy depends only on a few moment densities,
defined to be $\myphi_n=\sum_{i=1}^N f_n(\sigma_i)/V$, where $V$ is the
system volume and the functions $f_n(\sigma)$ are arbitrary. The sum is
over all particles in a homogeneous phase, and the $\sigma_i$ are
individual particle properties, such as size. If we set $f_0(\sigma)=1$
then we include the number density $\rho=\myphi_0$ amongst the moment
densities (in doing this though the $n=0$ term should be omitted from
certain sums below). As previously reported, moment densities can be
treated as independent thermodynamic density variables provided
the ideal part of the free energy density is suitably generalised:
\begin{equation}
\fid=\rho\kT(\ln\rho-1)-\rho T s(m_n)
\end{equation}
where $s(m_n)$ is a generalised entropy of mixing per particle, and
$m_n=\myphi_n/\rho=\sum_{i=1}^N f_n(\sigma_i)/N$ ($n>0$) are
generalised moments.  An expression for $s(m_n)$ can be given only
if the parent or feedstock distribution, $p(\sigma)$, is specified.
This is because one must know the parent distribution from which
particles are drawn, to conclude anything about the distributions in
daughter phases.  The generalised entropy of mixing is then given by a
Legendre transform (I will set $\kB=T=1$ in the following for
simplicity):
\begin{equation}
s(m_n)=h(\theta_n)+\sum_{n>0}\theta_nm_n,\qquad
m_n=-\frac{\partial h}{\partial\theta_n},\label{legeq}
\end{equation}
where the function $h$ is a generalised cumulant generating 
function for $p(\sigma)$,
\begin{equation}
h=\ln\int d\sigma\,p(\sigma)\,\exp\Bigl[-\sum_{n>0}\theta_n
f_n(\sigma)\Bigr].\label{heq}
\end{equation}
The excess free energy density, $\fex(\myphi_n)$, is added to
$\fid$ to obtain the overall free energy density $f$.  The spinodal
stability condition is then the standard one of positive definiteness
of the matrix of second partial derivatives of $f$ with respect to the
$\myphi_n$ (including $\myphi_0=\rho$ amongst these). For the ideal
part these second derivatives are readily shown to be
\begin{eqnarray}
&&\frac{\partial^2\!\fid}{\partial\myphi_0^2}=\frac{1}{\rho}
-\frac{1}{\rho}\sum_{n,r>0} m_n m_r
\frac{\partial^2s}{\partial m_n\partial m_r},\qquad
\frac{\partial^2\!\fid}{\partial\myphi_0\partial\myphi_n}
=\frac{1}{\rho}\sum_{r>0}
m_r\frac{\partial^2s}{\partial m_n\partial m_r},\quad(n>0),\nonumber\\
&&\hspace{10em}\frac{\partial^2\!\fid}{\partial\myphi_n\partial\myphi_r}
=-\frac{1}{\rho}\frac{\partial^2s}{\partial m_n\partial m_r},
\quad(n,r>0).\label{d2fid}
\end{eqnarray}
The key is clearly the matrix $\partial^2s/\partial m_n\partial m_r$
It can be easily shown from eq.~\eqref{legeq} that if $(\A)_{nr} =
\partial^2h/\partial\theta_n\partial\theta_r$ ($n,r>0$), then
$\partial^2s/\partial m_n\partial m_r = -(\A^{-1})_{nr}$.

Since the parent distribution enters into the solution in an essential
way, it constrains the mean values of the moments to those of the
parent distribution.  This constraint is essential to obtain the exact
spinodal and related stability conditions, and cloud and shadow
curves.  For a homogeneous system, it means that after all calculations
have been completed the results should be constrained to lie on the
physical dilution line, $\myphi_n/\rho = m_n =
\pav{f_n(\sigma)}$ where $\pav{\ldots}=\int
d\sigma\,p(\sigma)(\ldots)$.  Fortunately this constraint has a simple
implementation.  By explicit differentiation of eq.~\eqref{heq} the
moments are given as functions of $\theta_n$ by
\begin{equation}
m_n=e^{-h}\int d\sigma\,p(\sigma)f_n(\sigma)
\exp\Bigl[-\sum_{r>0}\theta_r f_r(\sigma)\Bigr].
\end{equation}
Since the map $\theta_n\leftrightarrow m_n$ should be $1:1$, otherwise
the Legendre transform is ill defined, the dilution line
constraint is seen to correspond to the point $\theta_n=0$.  This
observation now makes it obvious why stability conditions
such as the spinodal condition, the critical point, and so on depend
only on a finite set of moments or cumulants of the parent
distribution.  These conditions involve second or higher order
derivatives of $h$ evaluated on the dilution line, which corresponds
to the point $\theta_n=0$. Since $h$ is a cumulant generating
function, these derivatives are precisely the cumulants of
$p(\sigma)$.  This observation generalises a number of moment
truncation theorems reported previously by other workers \cite{Moms}.
Applying this dilution line constraint to the matrix $\A$ obtains for
instance
\begin{equation}
(\A)_{nr}=\langle f_n(\sigma)f_r(\sigma)\rangle_p
-\langle f_n(\sigma)\rangle_p\langle f_r(\sigma)\rangle_p.
\end{equation}
This matrix is readily inverted for use in eqs.~\eqref{d2fid}.

The spinodal stability limit corresponds to a vanishing eigenvalue of
the matrix of second partial derivatives of the free energy. The
associated eigenvector contains valuable but oft-neglected information
on the spinodal instability direction. This is the direction in
composition space in which the homogeneous system starts to become
unstable towards small perturbations, as the spinodal line is crossed.
It can be used to determine (mean-field) critical points, as points
where the instability direction is tangent to the spinodal line. Away
from such points, the instability direction tends to reflect the slope
of tielines connecting coexisting phases.

In the Sollich-Cates picture \cite{SC}, the spinodal instability is
contained in the family of size distributions
$p(\sigma)\exp[\sum\lambda_n f_n(\sigma)]$. Expanding this about a point
on the dilution line indicates that, in terms of a size distribution,
the spinodal instability direction is contained in
$\Delta\rho(\sigma)=\rho p(\sigma)\sum\Delta\lambda_nf_n(\sigma)$, where
the prefactor $\rho$ comes from the $n=0$ term. The $\Delta\lambda_n$
can be found from the constraints $\int d\sigma\,
f_n(\sigma)\Delta\rho(\sigma)=\Delta\myphi_n$ where $\Delta\myphi_n$ is
the prescribed eigenvector. From this, if
$(\B)_{nr}=\pav{f_n(\sigma)f_r(\sigma)}$ ($n,r\ge0$), the spinodal
instability direction is
\begin{equation}
\Delta\rho(\sigma)=p(\sigma)
\sum_{n,r\ge0}f_n(\sigma)(\B^{-1})_{nr}\Delta\myphi_r.\label{spineq}
\end{equation}
Actually, in the Sollich-Cates picture one can also show that the matrix
$\partial^2\!\fid / \partial\myphi_n \partial\myphi_r=
((\rho\B)^{-1})_{nr}$ \cite{Soll}. The equivalence to
eqs.~\eqref{d2fid} can be established with some algebra. Note that if
the $f_n(\sigma)$ ($n\ge0$) form an orthonormal set with $p(\sigma)$ as
a weight function, the matrices $\A$ and $\B$ become unit matrices, and
many of these results simplify markedly.

I now apply this machinery to examine the spinodal stability
of hard spheres.  It is convenient to introduce a fiducial diameter
$\sigma_0$ and scale all densities with $\pi\sigma_0^3/6$.
The excess free energy density corresponding to the BMCSL EOS
is given by Salacuse and Stell \cite{SS}
\begin{equation}
\frac{\pi\sigma_0^3}{6}\fex=
\Bigl(\frac{\myphi_2^3}{\myphi_3^2}-\myphi_0\Bigr)\ln(1-\myphi_3)
+\frac{3\myphi_1\myphi_2}{1-\myphi_3}
+\frac{\myphi_2^3}{\myphi_3(1-\myphi_3)^2}.
\end{equation}
The moment densities are $\myphi_n=(\pi\sigma_0^3/6)\rho m_n$ and the
moment functions themselves are $f_n(\sigma)=(\sigma/\sigma_0)^n$. Thus
$\myphi_0=\pi\sigma_0^3\rho/6$ is now a dimensionless number density, and
$\myphi_3=\phi$ is the volume fraction.

For the parent distribution I take either a Schulz distribution with
an extra monodisperse component or a mixture of two Schulz
distributions.  I will characterise these by the mean diameter and the
degree of polydispersity, where the latter is the ratio of the
standard deviation to the mean diameter (usually expressed as a
percentage).  The moments $\pav{(\sigma/\sigma_0)^n}$ which appear
in the matrices $\A$ and $\B$ are given by simple algebraic
relations once the distributions have been specified.

The $4\times4$ spinodal determinant is constructed as described above,
and examined as a function of the ratio of mean diameters,
$\sigma_L/\sigma_S$, the degree of polydispersity of the two
components, $\delta_L$ and $\delta_S$, and the volume fractions of the
two components, $\phi_L$ and $\phi_S$.  Generally the
determinant is always positive except when a large value of the diameter
ratio, $\sigma_L/\sigma_S\agt60$, is combined with a large degree of
polydispersity of the large spheres, $\delta_L\agt50$\%.

Fig.~1(a) shows the spinodal instability appearing as the diameter ratio
is increased, for a bimodal distribution with a monodisperse small
component. Fig.~1(b) shows the same for fixed mean diameter ratio and
increasing degree of polydispersity of both components. Interestingly,
at fixed mean diameter ratio and degree of polydispersity of the large
component, the instability region diminishes as the degree of
polydispersity of the small species increases. This is illustrated in
fig.~2. In both figures, results are only shown for the physically
reasonable region $\phi_S+\phi_L\alt0.5$.

The tick marks (short lines) on the spinodal lines indicate the spinodal
instability direction projected into the $(\phi_S,\phi_L)$ plane. The
insets to these figures show unnormalised parent distributions at one
selected point in each plot together with the spinodal instability
direction from eq.~\eqref{spineq}. Thus in fig.~1(b) for example, at the
marked point, the system is unstable towards a composition fluctuation
in which the density of large spheres increases and the density of small
spheres decreases: the precise functional form of the composition
fluctuation is the dashed line in the inset.

The tick marks and insets show that the instability is one of demixing
with some additional size partitioning. The instability direction, as
indicated by the tick marks for instance, is never found to be
tangential to the spinodal line in the physical region although it does
approach this condition for high values of the large sphere volume
fraction. This clearly points to a critical point lying at large
$\phi_L$ and suggests that the equilibrium state is one of a highly
dense fluid of large spheres coexisting with much less dense fluid of
mostly small spheres. Note that the degree of polydispersity for the
large spheres is easily large enough to suppress the formation of an
ordered phase \cite{Bart1}. 

Why should polydispersity have this effect? If one introduces small
spheres into a fluid of large spheres, depletion forces in the latter
are apparently insufficient to bring about phase separation, at least
according to the BMCSL EOS. It is known though that polydispersity
decreases the pressure of a hard sphere fluid \cite{BMCSL}, due to
relaxed packing constraints. It appears from the present calculation
that polydispersity can make the large sphere fluid sufficiently
compressible for the small sphere depletion forces to induce fluid-fluid
phase separation. What is remarkable is that the BMCSL EOS appears to
capture this effect, without being specifically designed to do so.

These results show intriguing parallels with the experimental
observations, with regard to the nature of phase separation in hard
sphere mixtures. In the experiments \cite{Expts,SMS,DYP}, phase
separation is observed at a much smaller diameter ratio, typically
$\sigma_L/\sigma_S\sim10$, and degree of polydispersity, typically
$\alt10$\%. The dense phase of large spheres is also observed to be
crystalline or amorphous. These differences are presumably due to the
sensitivity of the phase boundary to the accuracy of the EOS, as
reported by Biben and Hansen \cite{BH} (given this sensitivity, one can
speculate whether the discrepancies between experimental results
\cite{SMS,DYP} for size ratio dependence may be due to polydispersity
effects). Also, if the polydispersity in the dense phase is sufficiently
small, ordering into a crystal phase is likely to take place. The
novelty of the present analysis is that the broad picture is reproduced
by the BMCSL EOS, if polydispersity is taken into account.

Some intriguing possibilities still remain though for the true
equilibrium phase diagram of an arbitrarily polydisperse mixture. One
such is fractionation into multiple crystalline phases \cite{Bart2}.
Another is to relax the packing constraints in a dense crystal by
degrading the crystal symmetry from FCC, or the formation of
superlattice structures which could be regarded as interpenetrating
multiple crystals. At high volume fractions though, the glass transition
and other kinetic considerations undoubtably play a major role in
determining what is observed experimentally.

\stars

I acknowledge useful correspondence with {\sc Jos\'e Cuesta}, {\sc
Richard Sear} and {\sc Peter Sollich}.

\newpage

%
%

\begin{figure}
\centering
\includegraphics[width=2.5in]{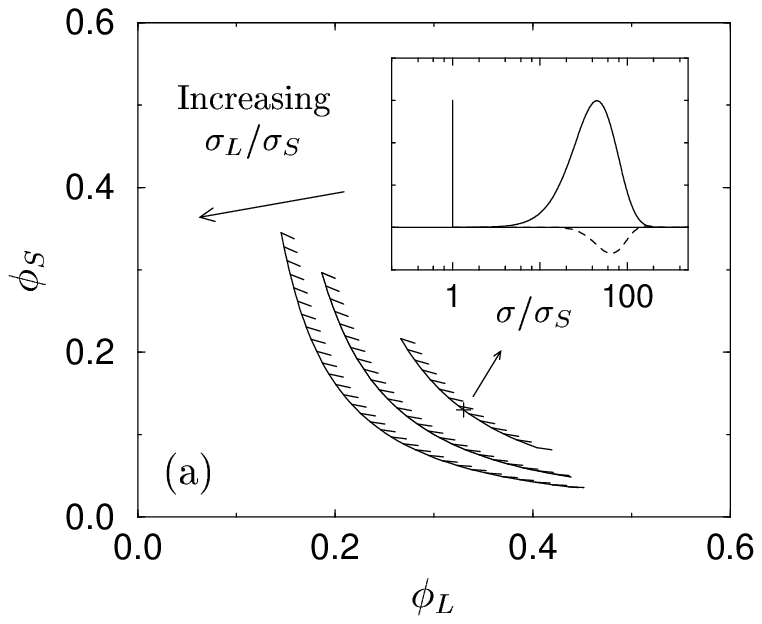}~%
\includegraphics[width=2.5in]{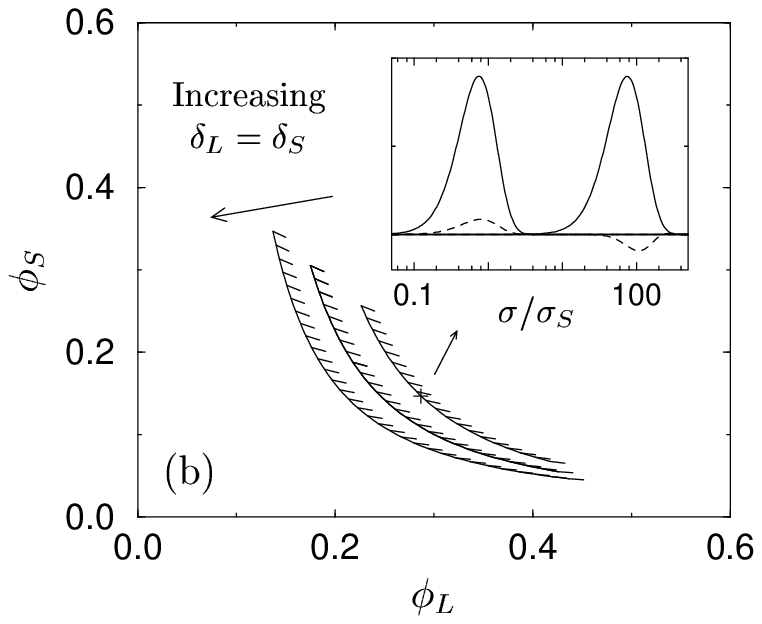}
\caption[?]{Spinodal instability lines for polydisperse hard sphere
mixtures (solid lines with tick marks in main plots): (a) monodisperse
small spheres mixed with polydisperse large spheres, $\delta_L=50$\%,
for $\sigma_L/\sigma_S=60$, 80 and 100, and (b) mixture of polydisperse
small and large spheres at $\sigma_L/\sigma_S=100$, for
$\delta_L=\delta_S=$50\%, 65\% and 90\%. The tick marks indicate the
spinodal instability direction. The insets show the (unnormalised) size
distribution (solid lines) and the direction in which it is spinodally
unstable (dotted lines), at the marked points.}
\end{figure}

\begin{figure}
\centering
\includegraphics[width=2.5in]{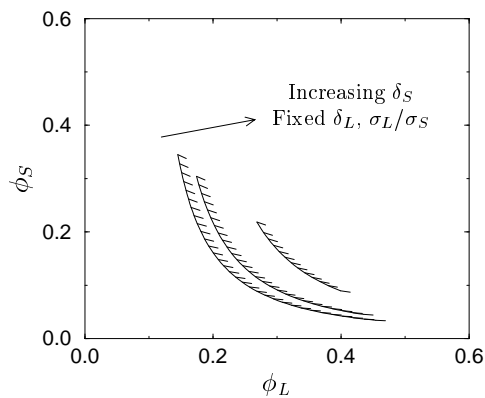}
\caption[?]{Mixture of polydisperse small and large spheres at
$\sigma_L/\sigma_S=100$, $\delta_L=50$\%, for $\delta_S=0$
(monodisperse), 30\% and 60\%.}
\end{figure}

\end{document}